\documentclass[
    ,final            
  ]
  {aipproc}

\layoutstyle{8x11single}

\begin{document}

\title{Measurement of the $\nu_{\mu}$-CCQE cross section in the SciBooNE experiment}

\classification{13.15.+g, 25.30.Pt}
\keywords{SciBooNE, absolute charged current quasi-elastic cross section}

\author{Jose Luis Alcaraz-Aunion}{
  address={IFAE-Universidad Autonoma de Barcelona, Spain}
}

\author{Joseph Walding}{
  address={Imperial College, London, SW7 2AZ, United Kingdom}
}

\begin{abstract}

SciBooNE is a neutrino and anti-neutrino cross-section experiment at Fermilab, USA.  The SciBooNE experiment is summarised and two independent CCQE analyses are described. For one of the analyses, an absolute $\nu_{\mu}$-CCQE cross section in the neutrino energy region (0.6-1.6) GeV is shown and the technique developed for such a purpose is also explained. The total cross section measured over this energy range agrees well with expectations, based on the NEUT event generator and using a value of 1.21 GeV for the CCQE axial mass.
\end{abstract}

\maketitle


\section{The SciBooNE Experiment}

\indent SciBooNE is a $\nu_{\mu}$ and  $\bar{\nu}_{\mu}$ scattering experiment based at the Fermi National Accelerator Laboratory, USA.  The main goal is to accurately measure the $\nu_{\mu}$ and $\bar{\nu}_{\mu}$ cross-sections on nuclear targets at energies around 1 GeV.  Currently few measurements have been made in this region on a nuclear target with high statistics.  These measurements are of interest to the next generation of oscillation experiments.
  
\subsection{The Booster Neutrino Beam-line}

Protons are accelerated up to 8 GeV and focused onto a 71 cm long, 1 cm wide Beryllium target.  The resultant meson shower (mostly pions) is then focused by a magnetic horn.  The horn has reversible polarity allowing us to sign select the mesons that enter the 50 m decay pipe downstream from the target, this allows us to run in neutrino or anti-neutrino mode.  The resultant mesons decay in flight producing our neutrino beam.  The short decay pipe minimizes the chance of secondary decays resulting in an intrinsic  $\nu_{e}$ background, important to oscillation studies.

\subsection{The SciBooNE Detector}
 The SciBooNE detector is composed of three sub-detectors: the SciBar detector, acting as a target,  the electromagnetic calorimeter (EC) and the muon range detector (MRD).\\

\indent The SciBar detector consists of 14,336 1.3 $\times$ 2.5 $\times$ 300.0 cm scintillator channels arranged in 64 layers comprised of one horizontal and one vertical plane read out by multi-Anode PMTs (MAPMT), with 64 channels read out per MAPMT.  The SciBar detector was originally used by the K2K collaboration as a part of a near detector before being dis-assembled and shipped to Fermilab where it was reconstructed for use by SciBooNE.\\

\indent The EC is the central detector sandwiched between SciBar and the MRD.  It consists of 2 planes - one horizontal, one vertical - each made up of 32 lead foil and scintillating fibre ``spaghetti'' calorimeter modules.  Because of this, the EC is 11 $X_{0}$ thick, able to measure large EM shower deposits exiting SciBar and thus identify $\pi^{0}$ and $\nu_{e}$ candidate events.  The EC was originally built for the CHORUS experiment before being used by K2K.\\

\indent The MRD is a new detector made up largely of recycled materials.  The detector consists of 362 scintillator paddles arranged into 13 alternating horizontal and vertical planes able to stop muons with momentum up to 1.2 GeV.  The role of the MRD is to range out and reconstruct muons resulting from Charged-Current interactions. 

\indent For the recycling effort and reuse of multiple detectors and electronic components the SciBooNE experiment was awarded a 2008 DOE P2* Environmental award.



\subsection{SciBooNE Operations}
\begin{figure}[ht]
\label{fig:SciBooNEPOT}
\includegraphics[width=0.35\textwidth]{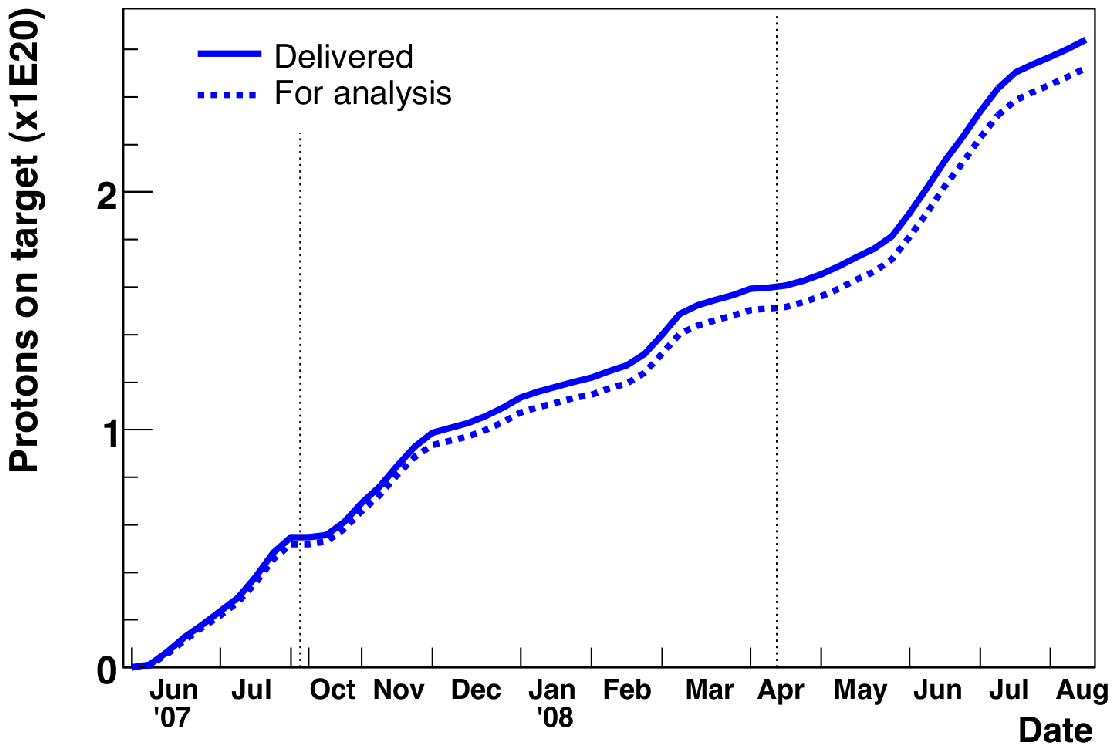}
\includegraphics[width=0.35\textwidth]{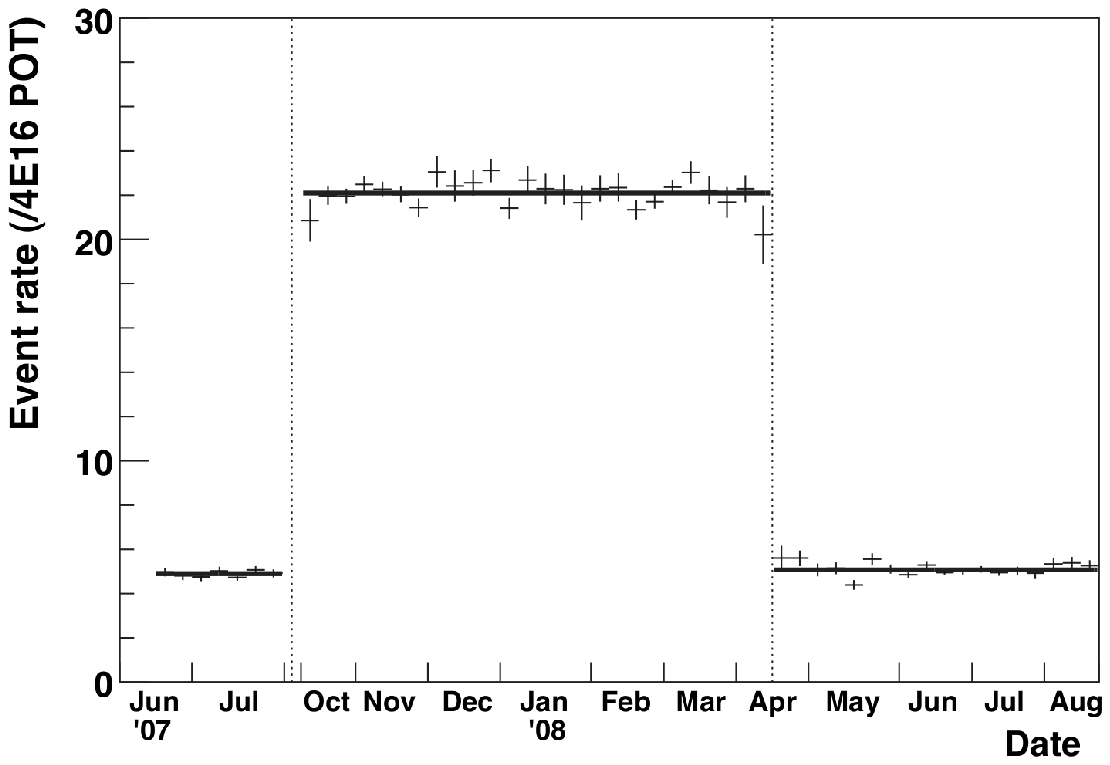}
\caption{The protons on target (POT) received during the full SciBooNE run (left) and the neutrino event rate per $4 \times 10^{16}$ POT (right).  The event rates during the two anti-neutrino runs are consistent.}

\end{figure}

SciBooNE ran from June 2007 until August 2008 collecting $0.99 \times 10^{20}$ Protons on Target (POT) in neutrino mode and $1.53 \times 10^{20}$ POT in anti-neutrino mode, 25\% more than requested.  The total POT and event rate for the entire run can be seen in figure \ref{fig:SciBooNEPOT}.

\subsection{SciBooNE MC Simulation}
The neutrino flux MC is estimated using extrapolated thin target data from the HARP experiment, with the errors estimated using the Sanford-Wang parameterisation \cite{MINIBOONE_flux}.
The NEUT library\cite{NEUT} is used to simulate the neutrino interactions with the nucleus in our MC simulation. Such a library is based on the following neutrino interaction models:  The Llewellyn-Smith Model is used to model the CCQE interactions with an $M_{A}$ value of 1.2 GeV/c$^{2}$.  The Rein-Sehgal model is used to model the CC1$\pi$ interactions, with the CC-coherent interactions suppressed by a factor of a third, in agreement with the SciBooNE measurement \cite{Hiraidecc}.  We use the Smith-Moniz Relativistic Fermi Gas nuclear model and Formation Zone Parameterisation for the final state interactions.  For carbon the Fermi momentum is set to 217 MeV and a binding energy of 27 MeV is applied.

\section{The CCQE Analyses}

\subsection{SciBar contained analysis}

\begin{figure}[ht]
\centering
\label{fig:MuonLifetime}
\includegraphics[width=0.42\textwidth]{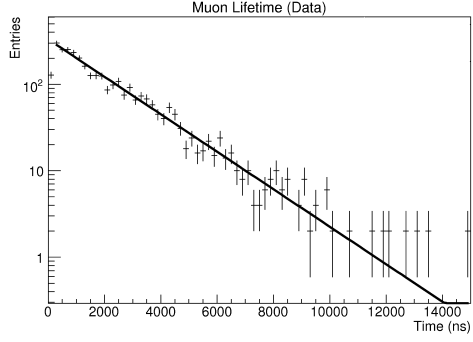}
\caption{Reconstructed muon lifetime in data.  Lifetime:  $2.003 \pm 0.047(stat)\mu s$ consistent with experiment \cite{Suzuki} }
\end{figure}

Pre-selection cuts are used to identify SciBar contained events.  The events must be completely contained within the SciBar fiducial volume ($-130 <x,y< 130$ and $2.62 < z < 157.2$ cm) and the earliest track in the event must be within the beam window of 2$\mu s$.  Charged-Current events are identified by tagging the decay electron signature from the stopped muon.  This is identified by looking for double-coincidences between horizontal and vertical SciBar TDC blocks.  Figure \ref{fig:MuonLifetime} shows the reconstructed muon lifetime in data for all identified events in SciBar.

\begin{figure}[ht]
\centering
\label{fig:ccqecut}
\includegraphics[width=0.42\textwidth]{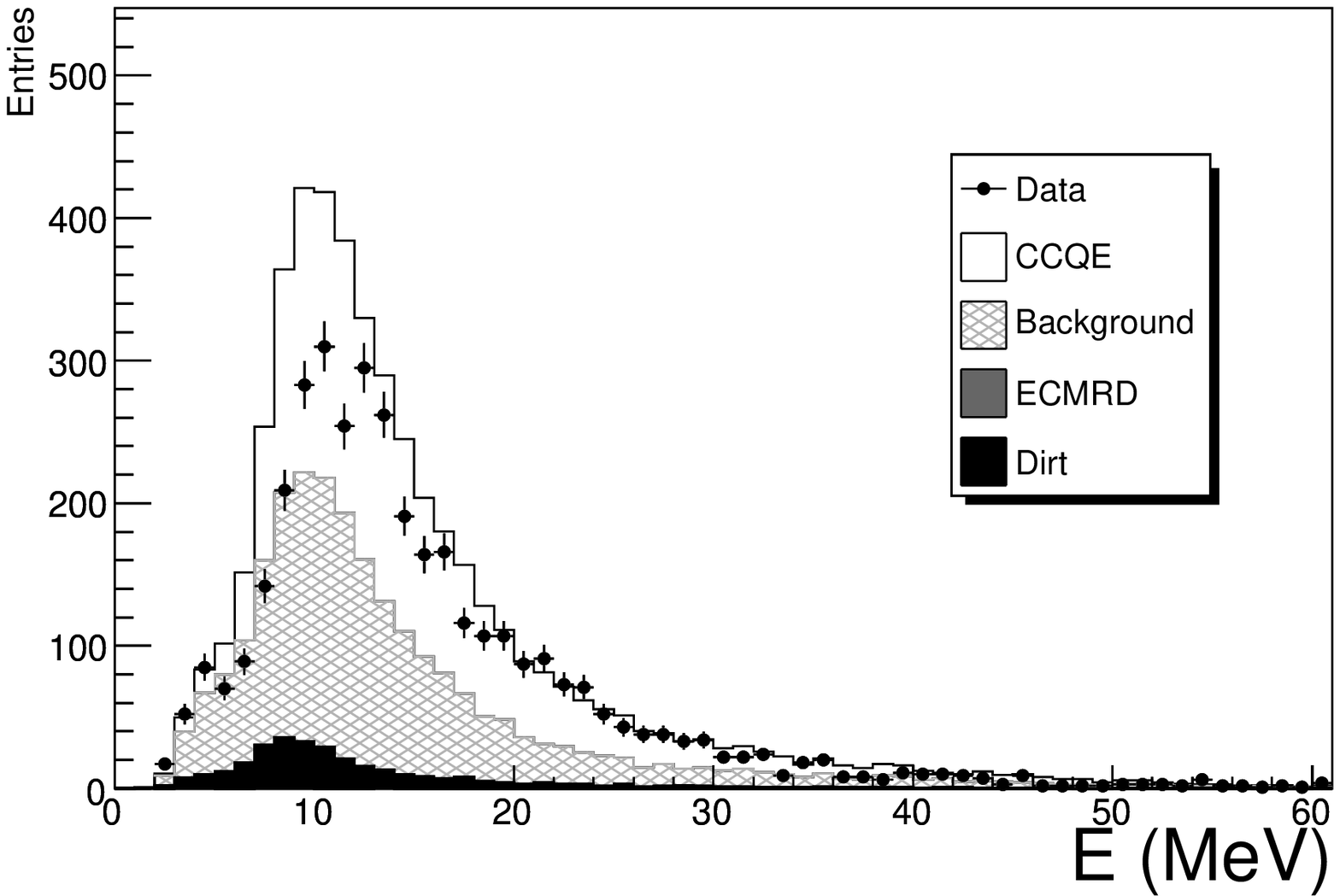}
\includegraphics[width=0.42\textwidth]{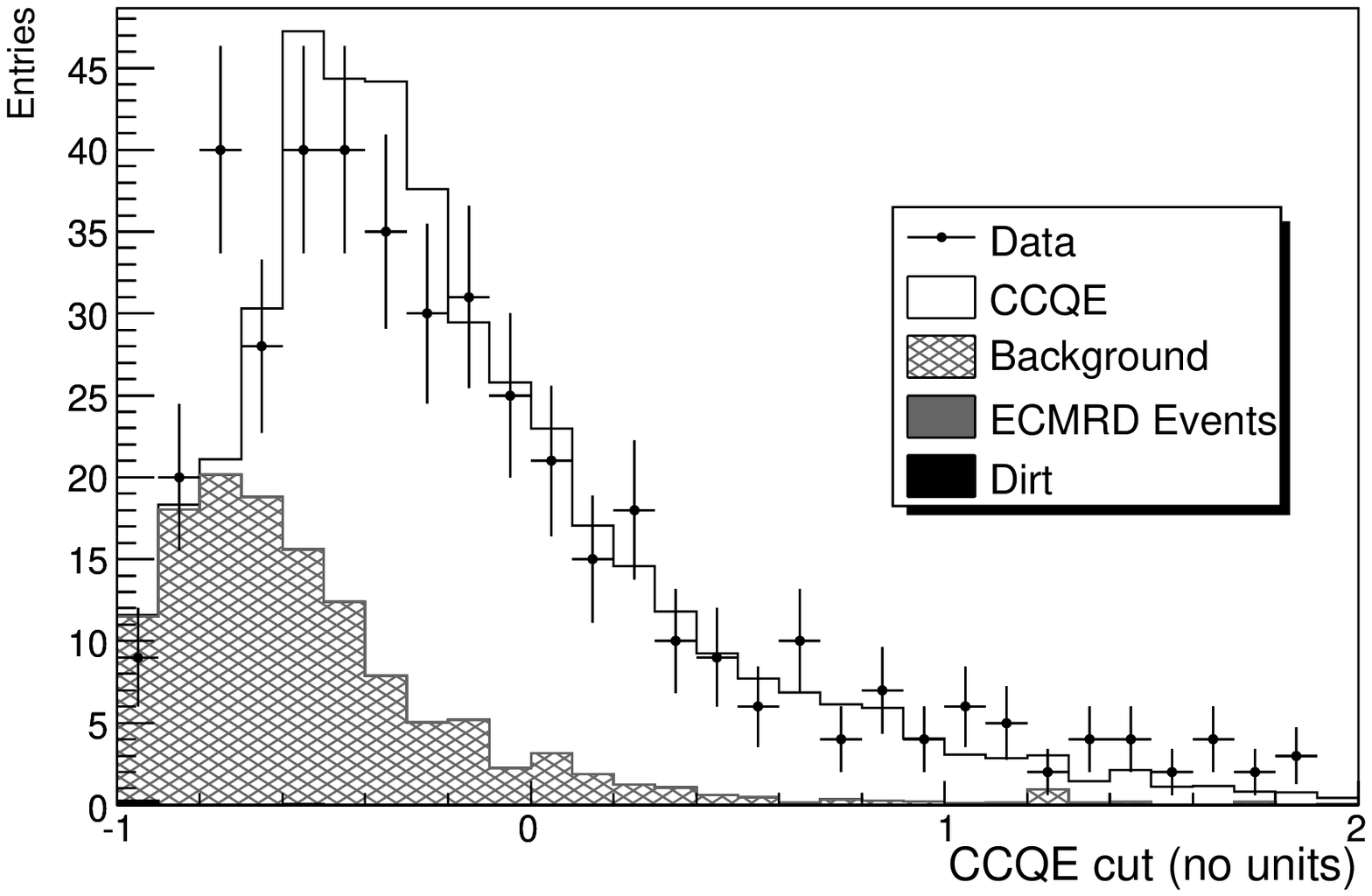}
\caption{From left to right:  The Vertex Activity cut applied to the 1-track sample, a cut of 8 MeV is applied.  The CCQE cut applied to the 2-track $\mu$ + p sample, a cut of -0.6 is applied.}
\end{figure}

To purify the CCQE event fraction further cuts are applied.  
For the 1-track sample a vertex activity cut of 8 MeV is applied to enhance the CCQE fraction with unreconstructed proton tracks, see figure \ref{fig:ccqecut}.
For multi-track events a Muon Confidence Level (MuCL) is constructed. Such a variable defines the probability of a particle being a MIP (or a muon-like track).  A cut of 0.05 is applied to the track with the lowest MuCL value (least muon-like track), identical to the SciBar-MRD analysis (see figure \ref{cuts}).  Those events that fail this cut become the 2-track $\mu + \pi$ sample used in the fitting architecture to constrain the background.
An additional CCQE cut is applied to the 2-track $\mu +$ p sample (see figure \ref{fig:ccqecut}) using the reconstructed neutrino energy assuming a CCQE interaction and the energy deposited in the detector.  The CCQE cut is defined as:
\begin{equation}
(E_{\nu}^{CCQE} - (T_{p} + E_{\mu})_{SciBar})/T_{p} = \epsilon
\end{equation}
Where $E_{\nu}^{CCQE}$ is the reconstructed neutrino energy assuming a CCQE interaction, $T_{p}$ and $E_{\mu}$ are the proton kinetic energy and the total muon energy reconstructed from the energy deposits in SciBar respectively, and $\epsilon$ is the unit-less parameter used for the cut.  For this analysis a cut of $\epsilon > -0.6$ is applied to the sample.

\begin{figure}[ht]
\centering
\label{fig:pmuthmu}
\includegraphics[width=0.5\textwidth]{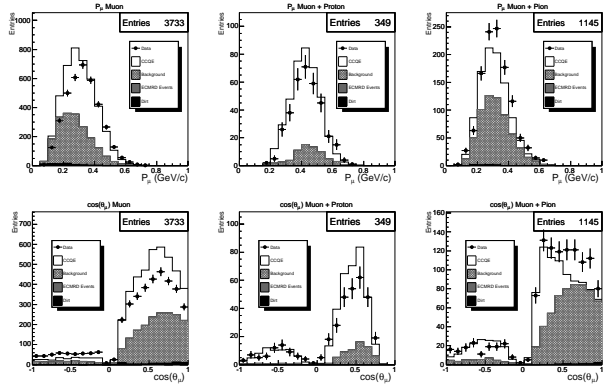}
\caption{P$_{\mu}$ (top) and $\theta_{\mu}$ (bottom) distributions (CCQE - white, background - hashed, dirt events - black) for all 3 final SciBar-contained samples, from left to right:  1-track $\mu$ events, 2-track $\mu + p$ events, 2-track $\mu + \pi$ events.  Dirt events occur when particles from a neutrino interaction outside the detector are detected within the detector.  This is a minor background in the SciBar-contained analysis and is negligible for the SciBar-MRD analysis. }
\end{figure}

The final sample selection is summarised in table \ref{table:SBContainedresults}. The 1-track sample is the largest statistical sample, with the 2-track ($\mu$ + p) having a high CCQE purity and the 2-track ($\mu$ + $\pi$) a high background purity.  Figure \ref{fig:pmuthmu} shows the P$_{\mu}$ and cos$(\theta_{\mu})$ distributions for all three samples.  The 2-track sample contains backwards tracks in data and MC, something under investigation.

\begin{table}
\caption{Summary table for the three SciBar contained sub-samples. }
\centering
\begin{tabular}{c c c c c}
\hline
\bf{Sample} & \bf{Data} & \bf{CCQE} & \bf{CC$\pi$} & \bf{Other}\\ [0.5ex]   
\hline                              
$\mu$         & 3733 & 55.5\% & 33.2\% & 11.3\% \\               
$\mu$ + p     & 349  & 86.3\% & 12.6\% & 1.1\%  \\
$\mu$ + $\pi$ & 1145 & 34.0\% & 47.9\% & 18.1\% \\ [1ex]         
\hline                              
\end{tabular}
\label{table:SBContainedresults}
\end{table}

\subsection{SciBar-MRD analysis}

Analogous to the previous analysis, the events in this analysis must satisfy two pre-selection cuts. The events must be produced within the beam time window ($0 < t < 2 \mu s$) and the interaction vertexes\footnote{The interaction vertex is defined as the upstream edge of the muon-track candidate} must be inside of the SciBar fiducial volume ($-130 <x,y< 130$ and $2.62 < z < 157.2$ cm).  The main difference to the SciBar-contained analysis is the requirement that the events contain at least one SciBar-track reaching the MRD detector (SciBar-MRD matched sample). Those tracks are defined as muon-track candidates. In addition, the muon tracks are required to stop inside of the MRD detector in order to reconstruct completely the neutrino energy, obtained under the assumption of a CCQE interaction. The resultant sample is called the SciBar-MRD stopped sample.

\begin{figure}[ht]
\includegraphics[height = 0.22\textwidth]{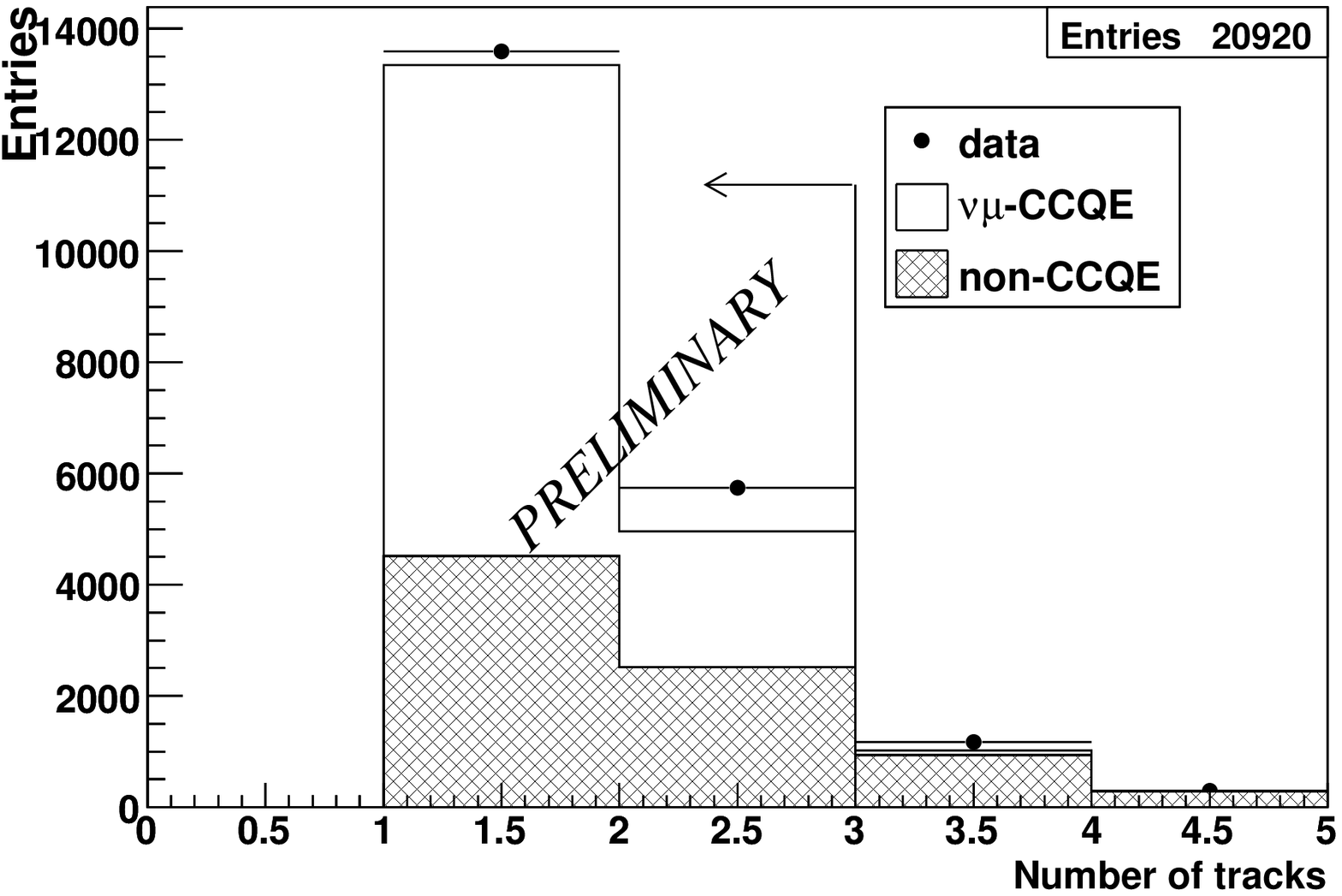}
\includegraphics[height = 0.22\textwidth]{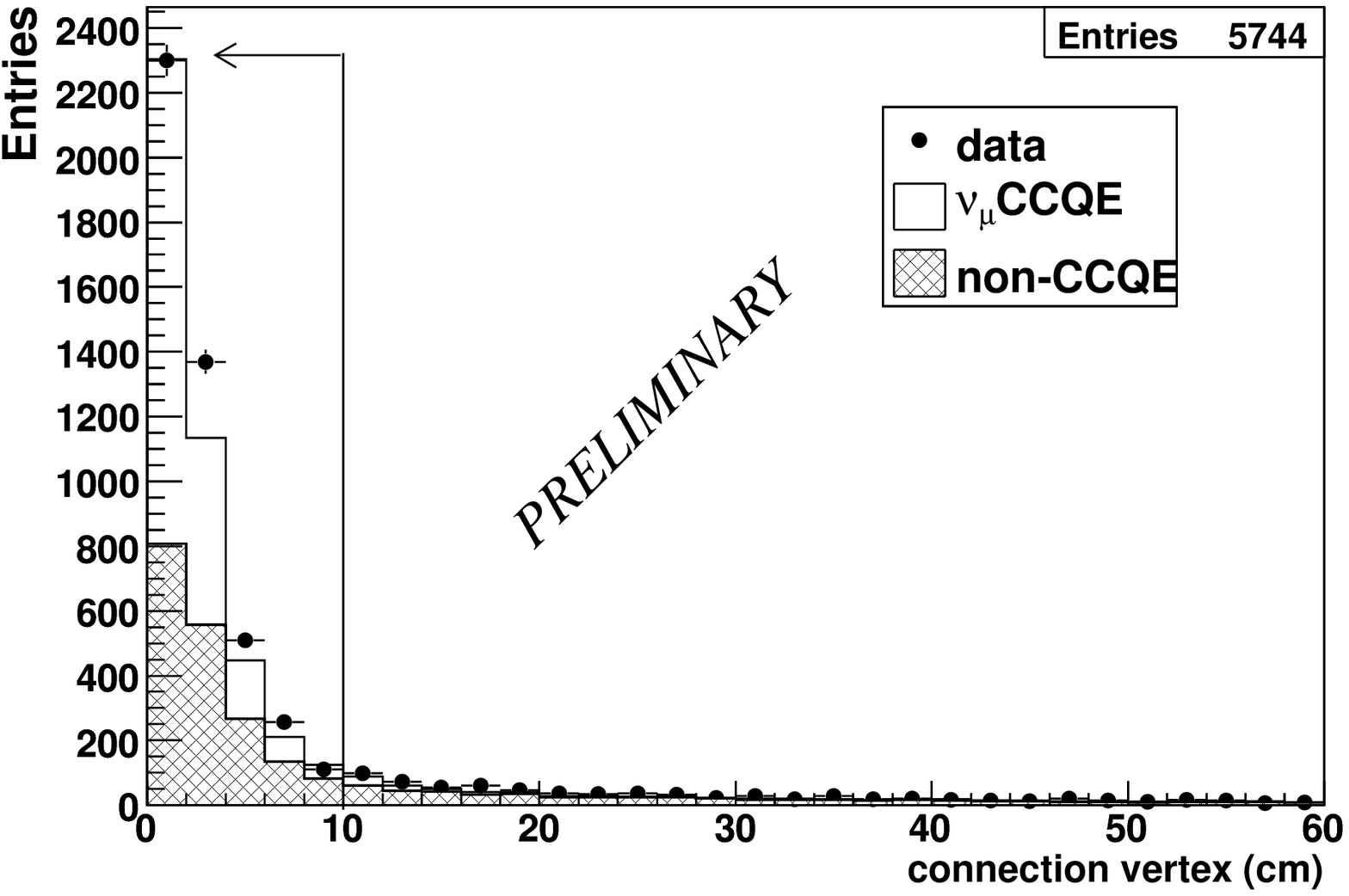}
\includegraphics[height = 0.22\textwidth]{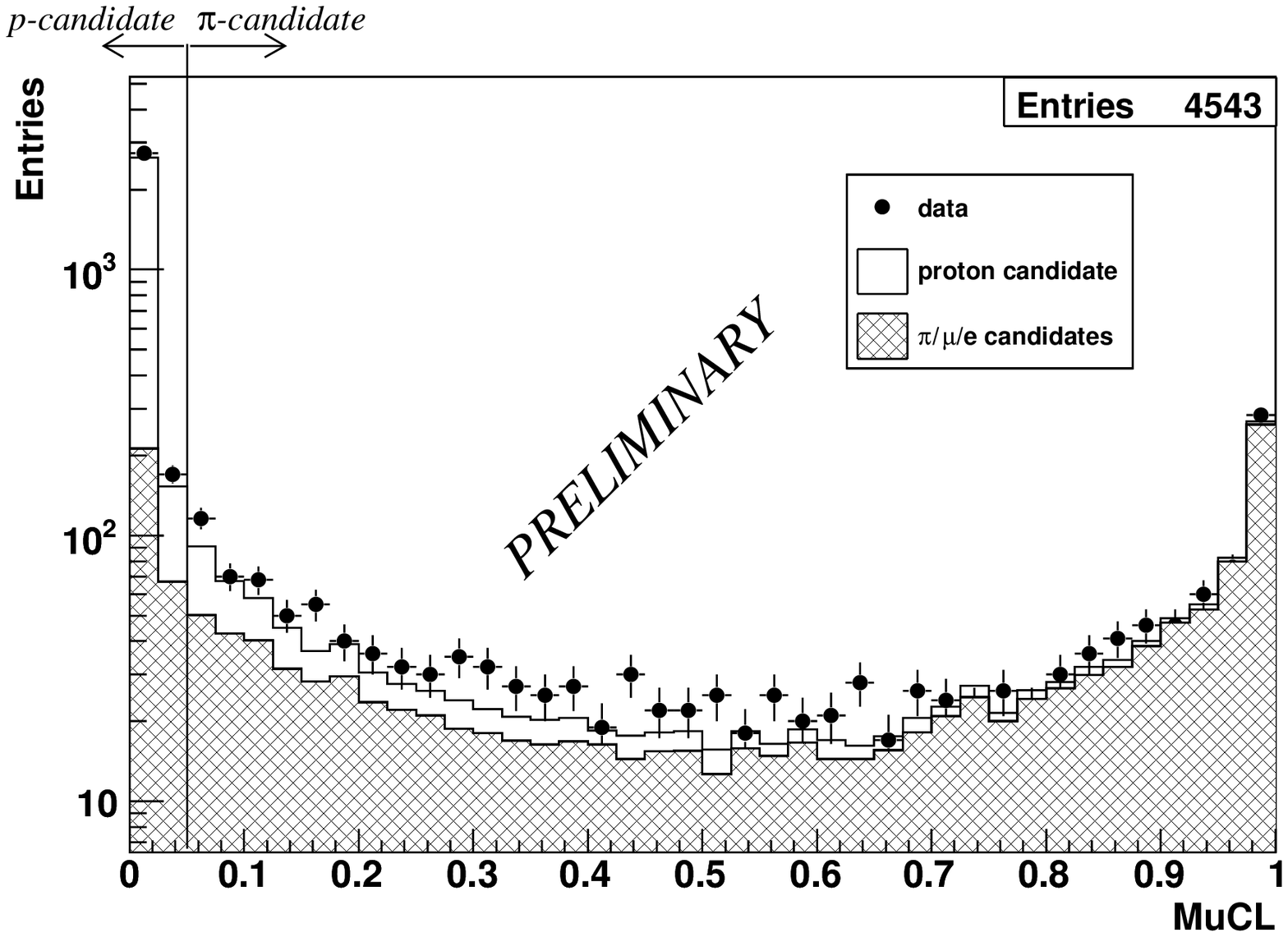}
\caption{From left to right, the distribution of number of tracks for the SciBar-MRD stop sample, the distribution of the connection vertex for the 2-track sample and the distribution of the MuCL variable for the 2-track sample after passing the connection vertex cut. The Monte Carlo is normalised to the SciBar-MRD matched sample.}
\label{cuts}
\end{figure}

The CCQE interactions are characterised by the low track multiplicity, the track vertex connection and the presence of a proton track. These three properties determine the CCQE cut selection. The track multiplicity cut (fig. \ref{cuts}) takes advantage of the QE event topology, selecting events with two tracks. However, the 1-track events are selected as well in order to include the QE events with low energy proton-tracks not reconstructed (which correspond with a proton momentum threshold of around 450 MeV/c).

The connection vertex and the particle identification cuts are based on properties of the second track, then both are applied to the 2-track sample. The connection vertex is defined as the distance between the neutrino interaction vertex, defined above, and the upstream edge of the second track. For QE interactions,  this distance must be close to zero. The figure \ref{cuts} shows the connection vertex distribution for the 2-track sample with cut value of 10 cm. This value has been selected by maximizing the efficiency times purity of the cut. 

The particle identification cut classifies the second track as proton-like or pion-like track according to a deposition energy study. For such a purpose, the MuCL variable (defined in the previous section) is used. Figure \ref{cuts} shows the MuCL distribution of the second track coming from 2-track events which pass the connection vertex cut. This figure shows that second tracks with MuCL values less than 0.05 are defined as proton-candidates while higher values correspond to pion-candidates.

The final selection is composed of three samples (see table \ref{summary_table}). The 1-track and 2-track  ($\mu$ + p) samples define the QE part of the selection and are characterised by a high statistics and high CCQE purity respectively. The 2-track ($\mu$ + $\pi$) sample defines the nonQE-part of the selection. This sample,  mainly composed of CC resonant $\pi^{+}$ production, is used in the fit to constrain the overall amount of CC nonQE background interactions with data, as explained below in the fit method description section.

\begin{table}
\begin{tabular}{rrccc}
\hline
    \tablehead{1}{r}{b}{Sample}
  & \tablehead{1}{r}{b}{data\\events}
  & \tablehead{1}{c}{b}{CCQE\\Efficiency (\%)\tablenote{calculated with respect to the number of CCQE events in the SciBar-MRD matched sample}}
  & \tablehead{1}{c}{b}{CCQE\\Purity (\%)} 
  & \tablehead{1}{c}{b}{ratio\\data/mc\tablenote{The monte carlo is normalized to the SciBar-MRD matched sample.}}     \\ 
\hline
1 track                   &  13586   &   52.9 & 65.2   & 1.03\\
2 track ($\mu$ + p)       &   2915   &   11.1 & 68.5   & 1.05\\
2 track ($\mu$ + $\pi$)   &   1628   &   2.7  & 32.3   & 1.14\\
\hline
\end{tabular}
\caption{Summary table with the three sub-samples selected for the extraction of the CCQE cross section.}
\label{summary_table}
\end{table}

\section{Extraction of absolute CCQE cross section}

\subsection{Fit method description}

The fit method is based on minimizing a likelihood function with the assumption that observed events follow a Poisson distribution. The likelihood expression is then written as follows \cite{pdg}:

\begin{equation}
\chi^{2}=2\times\sum_{i}\sum_{j}N^{exp}_{ij} -N^{obs}_{ij} +N^{obs}_{ij}\times ln\left(\frac{N^{exp}_{ij}}{N^{obs}_{ij}}\right), \label{likelihood}
\end{equation}
where $N^{exp}_{ij}$ and $N^{obs}_{ij}$ correspond to the expected and the observed number of events represented in bins of reconstructed muon momentum ($p^{\mu}_{i}$) and reconstructed muon angle ($\theta^{\mu}_{j}$). The number of expected MC events is mathematically expressed as follows:  

\begin{equation}
 N_{ij}^{exp}= \sum_{k} N^{QE,k}_{ij} + N^{nonQE}_{ij},
\label{mc_broken}
 \end{equation}
that is, the MC is broken into a QE component ($N^{QE,k}_{ij}$) which also depends of the true energy bins (given by the index \emph{k}) and a non-QE component ($N^{nonQE}_{ij}$). By taking expression \eqref{mc_broken}, the fit parameters are introduced in the following way:

\begin{equation}
 N_{ij}^{exp}= F_{N} 
        \left[ \sum_{k}^{n=10} a_{k} N^{QE,k}_{ij}  
        + a_{bck} N^{nonQE}_{ij} \right].
\label{fit_parameters}
 \end{equation}

A total of twelve parameters define the fit, two of them are global parameters ($F_{N}$ and $a_{bck}$) and the other ten are neutrino energy dependent parameters (${a_{1},a_{2}.....,a_{10}}$). $F_{N}$ is a scale factor, containing information about the data/MC normalization. \textbf{$a_{bck}$} gives information about the non-QE contamination. Finally, the energy dependent parameters ($a_{k}$), each one associated with a different true neutrino energy bin, return information of the neutrino energy shape. 


The absolute CCQE cross section is then evaluated for each neutrino energy bin using the following formula:

\begin{equation}
\sigma_{\nu_{\mu}}(E_{\nu}^{k}) = F_{N} \times a_{k} \times \sigma^{NEUT}_{\nu_{\mu}}(E_{\nu}^{k}),
\label{xsec_equation}
\end{equation}

where $\sigma^{NEUT}_{\nu_{\mu}}(E_{\nu}^{k})$ corresponds to the value of the absolute cross section predicted by NEUT in the neutrino energy bin \emph{k}. One can observe again the role of the $a_{k}$ parameters in the expression \eqref{xsec_equation}, re-weighting the predicted cross section for each neutrino energy bin.

\subsection{Results}

\begin{figure}[ht]
\includegraphics[height = 0.3\textwidth]{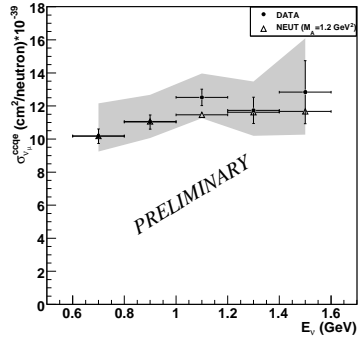}
\caption{The absolute $\nu_{\mu}$-CCQE cross section measured in SciBooNE for the SciBar-MRD analysis (circle dots) in comparison with the NEUT prediction (triangle dots). The statistics (bars) and the flux systematic errors (grey band) are included in the data.}
\label{final_xsection}
\end{figure}

The absolute $\nu_{\mu}$-CCQE cross section per nucleon shown in this proceeding (see fig. \ref{final_xsection}) is extracted only by using the sample selection from the SciBar-MRD analysis. The neutrino energy range of the cross section is then limited by the sensitivity of this selection, which only works in the interval (0.6-1.6) GeV. The low energy region will be covered by adding the SciBar-contained sample (work in progress). 

The cross section results are consistent with the theoretical prediction given by NEUT\cite{NEUT}, as figure \ref{final_xsection} shows.
 The systematic uncertainty is due to the limited knowledge in the neutrino flux predictions at SciBooNE.
 The dominant contribution to the flux systematics comes from the uncertainty in the rate and kinematics of $\pi^{+}$-production in the interactions of beam protons with the beryllium target, as measured by HARP \cite{HARP},\cite{MINIBOONE_flux}. Other flux systematics considered include uncertainties in other hadronic interactions, in beam proton delivery, and in horn magnetic focusing \cite{MINIBOONE_flux}.

\begin{theacknowledgments}

 The SciBooNE collaboration gratefully acknowledges support from various grants and contracts from the Department of Energy (U.S.), the National Science Foundation (U.S.), MEXT (Japan), INFN (Italy), the Science and Technology Facilities Council (U.K.), and the Spanish Ministry of Education and Science. 

\end{theacknowledgments}



\bibliographystyle{aipproc}   

\bibliography{nuint09_totalv1}





\end{document}